\documentclass[12pt]{article}

\input{tcilatex2.5}
\begin{document}

\title{The Exact Quantum Sine-Gordon Field Equation and Other Non-Perturbative
Results}
\author{H. Babujian\thanks{%
Permanent address: Yerevan Physics Institute, Alikhanian Brothers 2,
Yerevan, 375036 Armenia.} \thanks{
e-mail: babujian@lx2.yerphi.am, babujian@physik.fu-berlin.de}~ and M.
Karowski\thanks{
e-mail: karowski@physik.fu-berlin.de} \\
{\small \textit{Institut f\"ur Theoretische Physik}}\\
{\small \textit{Freie Universit\"at Berlin, Arnimallee 14, 14195 Berlin,
Germany}}}
\maketitle

\begin{abstract}
Using the methods of the ``form factor program'' exact expressions of all
matrix elements are obtained for several operators of the quantum sine
Gordon model: all powers of the fundamental bose field, general exponentials
of it, the energy momentum tensor and all higher currents. It is found that
the quantum sine-Gordon field equation holds with an exact relation between
the ``bare'' mass and the renormalized mass. Also a relation for the trace
of the energy momentum is obtained. The eigenvalue problem for all higher
conserved charges is solved. All results are compared with Feynman graph
expansions and full agreement is found.\\[8pt]
\end{abstract}

The classical sine-Gordon model is given by the wave equation
\[
\Box \varphi (t,x)+\frac{\alpha }{\beta }\sin \beta \varphi (t,x)=0.
\]
Since Coleman \cite{Co} found the wonderful duality between the quantum
sine-Gordon and the massive Thirring model a lot of effort has been made to
understand this quantum field theoretic model. A further contribution in
this direction is the present paper of which an extended version will be
published elsewhere \cite{BK}. The ``bootstrap'' program (see e.g. \cite{K2}%
) for integrable quantum field theoretical models in 1+1 dimensions starts
with the two particle sine-Gordon S-matrix for the scattering of fundamental
bosons (lowest breathers) \cite{KT}
\[
S(\theta )=\frac{\sinh \theta +i\sin \pi \nu }{\sinh \theta -i\sin \pi \nu }
\]
where $\theta $ is the rapidity difference defined by $p_{1}p_{2}=m^{2}\cosh
\theta $ and $\nu $ is related to the coupling constant by $\nu =\beta
^{2}/(8\pi -\beta ^{2}).$

From the S-matrix off-shell quantities as arbitrary matrix elements of local
operators are obtained by means of the ``form factor program'' \cite{KW}. In
particular we provide exact expressions for all matrix elements of all
powers of the fundamental bose field $\varphi (t,x)$ and its exponential $%
\mathcal{N}\exp i\gamma \varphi (t,x)$ for arbitrary $\gamma $. Here and in
the following $\mathcal{N}$ denotes normal ordering with respect to the
physical vacuum which means in particular for the vacuum expectation value $%
\langle \,0\,|\,\mathcal{N}\exp i\gamma \varphi (t,x)|\,0\,\rangle =1$. For
the exceptional value $\gamma =\beta $ we find that the operator $\Box ^{-1}%
\mathcal{N}\sin \beta \varphi (t,x)$ is local. Moreover the quantum
sine-Gordon field equation\footnote{%
In the framework of constructive quantum field theory quantum field
equations where considered in \cite{Sch,Fr}.}
\begin{equation}
\Box \varphi (t,x)+\frac{\alpha }{\beta }\mathcal{N}\sin \beta \varphi
(t,x)=0  \label{f}
\end{equation}
is fulfilled for all matrix elements, if the ``bare'' mass $\sqrt{\alpha }$
is related to the renormalized mass by\footnote{%
One of the authors (H.B.) has learned from Al.B. Zamolodchikov that this
mass relation is consistent with his results in \cite{Za}.}
\begin{equation}
\alpha =m^{2}\frac{\pi \nu }{\sin \pi \nu }  \label{mass}
\end{equation}
where $m$ is the physical mass of the fundamental boson. The factor $\frac{%
\pi \nu }{\sin \pi \nu }$ modifies the classical equation and has to be
considered as a quantum correction. For the sinh-Gordon model an analogous
quantum field equation has been obtained in \cite{MS}\footnote{%
It should be obtained from (\ref{f}) by the replacement $\beta \rightarrow
ig $. However the relation between the bare and the renormalised mass in
\cite{MS} differs from the analytic continuation of (\ref{mass}) by a factor
which is $1+O(\beta ^{4})\neq 1$.}. Note that in particular at the `free
fermion point' $\nu \rightarrow 1~(\beta ^{2}\rightarrow 4\pi )$ this factor
diverges, a phenomenon which is to be expected by short distance
investigations \cite{ST}. For fixed bare mass square $\alpha $ and $\nu
\rightarrow 2,3,4,\dots $ the physical mass goes to zero. These values of
the coupling are known to be specific: 1. the Bethe ansatz vacuum in the
language of the massive Thirring model shows phase transitions \cite{Ko} and
2. the model at these points is related \cite{K3,LeC,Sm1} to Baxters
RSOS-models which correspond to minimal conformal models with central charge
$c=1-6/(\nu (\nu +1))$ (see also \cite{MS}).

Also we calculate all matrix elements of all higher local currents $%
J_{M}^{\mu }(t,x)$ ($M=\pm 1,\pm 3,\dots $) fulfilling $\partial _{\mu
}J_{M}^{\mu }(t,x)=0$ which is characteristic for integrable models. The
higher charges fulfill the eigenvalue equation
\begin{equation}
\left( \int dxJ_{M}^{0}(x)-\sum_{i=1}^{n}\left( p_{i}^{+}\right) ^{M}\right)
|\,p_{1},\dots ,p_{n}\rangle ^{in}=0.  \label{J}
\end{equation}
In particular for $M=\pm 1$ the currents yield the energy momentum tensor $%
T^{\mu \nu }=T^{\nu \mu }$ with $\partial _{\mu }T^{\mu \nu }=0$. We find
that its trace fulfills
\begin{equation}
T_{~\mu }^{\mu }(t,x)=-2\frac{\alpha }{\beta ^{2}}\left( 1-\frac{\beta ^{2}}{%
8\pi }\right) \left( \mathcal{N}\cos \beta \varphi (t,x)-1\right) .
\label{T}
\end{equation}
This formula is consistent with renormalization group arguments \cite{Z,Ca}.
In particular this means that $\beta ^{2}/4\pi $ is the anomalous dimension
of $\cos \beta \varphi $. Again this operator equation is modified by a
quantum correction ($1-\beta ^{2}/8\pi $). Obviously for fixed bare mass
square $\alpha $ and $\beta ^{2}\rightarrow 8\pi $ the model will be
conformal invariant which is related to a Berezinski-Kosterlitz-Thouless
phase transition \cite{KT,JKKN,Wi}\footnote{%
This was pointed out to us by S.A. Bulgadaev.}. The proofs of the statements
(\ref{f}) -- (\ref{T}) is sketched in the following together with some
checks in perturbation theory. The complete proofs will be published
elsewhere \cite{BK}.

Let $\mathcal{O}(x)$ be a local scalar operator. The generalized form
factors are defined by the vacuum -- n-particle matrix elements
\[
\langle \,0\,|\,\mathcal{O}(x)\,|\,p_{1},\dots ,p_{n}\,\rangle
^{in}=e^{-ix(p_{1}+\dots +p_{n})}\,f_{n}^{\mathcal{O}}(\theta _{1},\dots
,\theta _{n})~,~~~\mathrm{for}~~\theta _{1}>\dots >\theta _{n}
\]
where the $\theta _{i}$ are the rapidities of the particles $p_{i}^{\mu
}=m(\cosh \theta _{i},\sinh \theta _{i})$. In the other sectors of the
variables the functions $\,f_{n}^{\mathcal{O}}(\theta _{1},\dots ,\theta
_{n})$ are given by analytic continuation with respect to the $\theta _{i}$.
General matrix elements are obtained from $f_{n}^{\mathcal{O}}(\underline{%
\theta })$ by crossing which means in particular analytic continuation $%
\theta _{i}\rightarrow \theta _{i}\pm i\pi $.

A form factor of n fundamental bosons (lowest breathers) is of the form \cite
{KW}
\[
f_{n}^{\mathcal{O}}(\underline{\theta })=N_{n}^{\mathcal{O}}K_{n}^{\mathcal{O%
}}(\underline{\theta })\prod_{1\leq i<j\leq n}F(\theta _{ij})
\]
where $N_{n}^{\mathcal{O}}$ is a normalization constant, $\theta
_{ij}=\theta _{i}-\theta _{j}$ and $F(\theta )$ is the two particle form
factor function. It fulfills Watson's equations
\[
F(\theta )=F(-\theta )S(\theta )=F(2\pi i-\theta )
\]
with the S-matrix given above. Explicitly it is given by the integral
representation \cite{KW}
\[
F(\theta )=N\exp \int_{0}^{\infty }\frac{dt}{t}\,\frac{\left( \cosh \frac{1}{%
2}t-\cosh (\frac{1}{2}+\nu )t\right) \left( 1-\cosh t(1-\frac{\theta }{i\pi }%
)\right) }{\cosh \frac{1}{2}t\sinh t}
\]
normalized such that $F(\infty )=1$. The K-function $K_{n}^{\mathcal{O}}(%
\underline{\theta })$ is meromorphic, symmetric and periodic (under $\theta
_{i}\rightarrow \theta _{i}+2\pi i$).

For an odd number of solitons and anti-solitons \cite{BFKZ} and for
chargeless operators \cite{BK} a general integral representation of
generalized form factors has been proposed. Using this integral
representations and the fusion `soliton + anti-soliton $\rightarrow $
breather' the following general formula for the K-function has been derived
in \cite{BK}\footnote{%
Bosonic sine-Gordon form factors have also been derived in \cite{Sm}. For
the sinh-Gordon model form factors where obtained in \cite{FMS,KM,MS} in
terms of determinants of symmetric polynomials. They are related to those of
this paper by the analytic continuation $\beta \rightarrow ig$.}. It turns
out to be of the form
\begin{equation}
K_{n}^{\mathcal{O}}(\underline{\theta })=\sum_{l_{1}=0}^{1}\dots
\sum_{l_{n}=0}^{1}(-1)^{l_{1}+\dots +l_{n}}\prod_{1\leq i<j\leq n}\left(
1+(l_{i}-l_{j})\frac{i\sin \pi \nu }{\sinh \theta _{ij}}\right) p_{n}^{%
\mathcal{O}}(\underline{\theta },\underline{z})  \label{K}
\end{equation}
where $z_{i}=\theta _{i}-\frac{i\pi }{2}\left( 1+(2l_{i}-1)\nu \right) $.
The dependence on the operator is encoded in the 'p-function' $p_{n}^{%
\mathcal{O}}$. It is separately symmetric with respect to the variables $%
\underline{\theta }$ and $\underline{z}$ and has to fulfill some simple
conditions in order that the form factor function $f_{n}^{\mathcal{O}}$
fulfill some properties \cite{KW,Sm}. These properties follow (see \cite
{BFKZ}) from general LSZ-assumptions and in additions specific features
typical for integrable field theories. In particular the recursion relation
holds
\begin{equation}
\limfunc{Res}_{\theta _{12}=i\pi }\,f_{n}^{\mathcal{O}}(\theta _{1},\dots
,\theta _{n})=2i\,f_{n-2}^{\mathcal{O}}(\theta _{3},\dots ,\theta
_{n})\left( \mathbf{1}-S(\theta _{2n})\dots S(\theta _{23})\right) .
\label{iii}
\end{equation}
Here we will not provide more details but only give some examples of
operators and their corresponding p-functions:

\begin{enumerate}
\item  The correspondence of exponentials of the field and their p-function%
\footnote{%
For the sinh-Gordon model an analogous representation as (\ref{K}) together
with this p-function was obtained in \cite{BL} by different methods.} is
\begin{equation}
\mathcal{N}e^{i\gamma \varphi }\leftrightarrow
\prod_{i=1}^{n}e^{(2l_{i}-1)i\pi \nu \gamma /\beta }  \label{q}
\end{equation}
for an arbitrary constant $\gamma $.

\item  Taking derivatives of this formula with respect to $\gamma $ we get
for the field and its powers
\begin{equation}
\mathcal{N}\varphi ^{N}\leftrightarrow \left(
\sum_{i=1}^{r}(2l_{i}-1)\right) ^{N}.  \label{N}
\end{equation}

\item  Higher currents (for $M=\pm 1,\pm 3,\dots $) correspond to the
p-functions
\begin{equation}
J_{M}^{\pm }\leftrightarrow \sum_{i=1}^{n}e^{\pm \theta
_{i}}\sum_{i=1}^{n}e^{Mz_{i}}  \label{M}
\end{equation}
for $n=$ even and zero for $n=$ odd. For $M=\pm 1$ we get the light cone
components of the energy momentum tensor $T^{\rho \sigma }=J_{\sigma }^{\rho
}$ with $\rho ,\sigma =\pm $ (see also \cite{MS}).
\end{enumerate}

In order to prove equations as for example (\ref{f}) and (\ref{T}) we
consider the corresponding p-functions and their K-functions defined by (\ref
{K}). The K-functions are rational functions of the $x_{i}=e^{\theta _{i}}$.
We analyze the poles and the asymptotic behavior and find identities by
induction and Liouville's theorem. Transforming these identities to the
corresponding form factors one finds the field equation (\ref{f}) and the
trace equation (\ref{T}) up to normalizations.

Normalization constants are obtained in the various cases by the following
observations:

\begin{enumerate}
\item[a)]  The normalization a field annihilating a one-particle state is
given by the vacuum one-particle matrix element, in particular for the
fundamental bose field one has
\[
\langle \,0\,|\,\varphi (0)\,|\,p\,\rangle =\sqrt{Z^{\varphi }}\quad \mathrm{%
with}\quad Z^{\varphi }=(1+\nu )\frac{\frac{\pi }{2}\nu }{\sin \frac{\pi }{2}%
\nu }\exp \left( -\frac{1}{\pi }\int_{0}^{\pi \nu }\frac{t}{\sin t}dt\right)
\]
where $Z^{\varphi }=1+O(\beta ^{4})$ is the finite wave function
renormalization constant calculated in \cite{KW}. This gives the
normalization constant
\begin{equation}
N_{1}^{(1)}=\sqrt{Z^{\varphi }}/2  \label{N1}
\end{equation}
for the form factors of the fundamental bose field and which are obtained
from the p-function of (\ref{N}) for $N=1$.

\item[b)]  If a local operator is connected to an observable like a charge $%
Q=\int dx\,\mathcal{O}(x)$ we use the relation
\[
\langle \,p^{\prime }\,|\,Q\,|\,p\,\rangle =q\langle \,p^{\prime
}\,|\,\,p\,\rangle .
\]
For example for the higher charges we obtain
\[
N_{2}^{J_{M}}=\frac{i^{M}m^{M+1}}{2\sin \pi \nu \sin \frac{M}{2}\pi \nu
F(i\pi )}\quad \mathrm{with}\quad \frac{1}{F(i\pi )}=Z^{\varphi }\frac{\beta
^{2}}{8\pi \nu }\frac{\sin \pi \nu }{\pi \nu }.
\]

\item[c)]  We use Weinberg's power counting theorem for bosonic Feynman
graphs \cite{BK}\footnote{%
This type of arguments has been also used in \cite{KW,FMS,KM,MS}.}. For the
exponentials of the boson field $\mathcal{O=N}e^{i\gamma \varphi }$ this
yields in particular the asymptotic behavior
\[
f_{n}^{\mathcal{O}}(\theta _{1,}\theta _{2,}\dots )=f_{1}^{\mathcal{O}%
}(\theta _{1})\,f_{n-1}^{\mathcal{O}}(\theta _{2,}\dots )+O(e^{-\func{Re}%
\theta _{1}})
\]
as $\func{Re}\theta _{1}\rightarrow \infty $ in any order of perturbation
theory. This behavior is also assumed to hold for the exact form factors.
Applying this formula iteratively we obtain from (\ref{K}) with (\ref{q})
the following relation for the normalization constants of the operators $%
\mathcal{N}e^{i\gamma \varphi }$%
\[
N_{n}^{\gamma }=\left( N_{1}^{\gamma }\right) ^{n}\quad (n\geq 1).
\]

\item[d)]  The recursion relation (\ref{iii}) relates $N_{n+2}$ and $N_{n}.$
For all p-functions mentioned above we obtain
\[
N_{n+2}=N_{n}\frac{2}{\sin \pi \nu F(i\pi )}\quad (n\geq 1).
\]
\end{enumerate}

Using c) and d) we calculate the normalization constants for the exponential
of the field $\mathcal{N}e^{i\gamma \varphi }$ and obtain
\begin{equation}
N_{1}^{\gamma }=\sqrt{Z^{\varphi }}\frac{\beta }{2\pi \nu }.  \label{Nq}
\end{equation}

The normalization constants (\ref{N1}) and (\ref{Nq}) now yield the field
equation (\ref{f}) with the mass relation (\ref{mass}). The statement (\ref
{T}) is proved similarly. The eigen value equation (\ref{J}) is obtained by
taking the scalar products with arbitrary out-states and by using a general
crossing formula \cite{BK}.

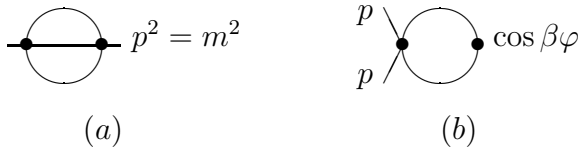
\begin{figure}[tbh]
\[
\begin{array}{l}
\unitlength5mm \begin{picture}(14,3) \put(1,2){\oval(2,2)}
\put(0,2){\makebox(0,0){$\bullet$}} \put(2,2){\makebox(0,0){$\bullet$}}
\put(-.5,2){\line(1,0){3}} \put(2.8,2){$p^2=m^2$} \put(1.5,-.5){$(a)$}
\put(11,2){\oval(2,2)} \put(10,2){\makebox(0,0){$\bullet$}}
\put(12,2){\makebox(0,0){$\bullet$}} \put(12.4,2){$\cos \beta \varphi$}
\put(10,2){\line(-1,2){.5}} \put(10,2){\line(-1,-2){.5}} \put(8.8,1){$p$}
\put(8.8,2.7){$p$} \put(11,-.5){$(b)$} \end{picture}
\end{array}
\]
\caption{Feynman graphs}
\label{f8}
\end{figure}
All the results may be checked in perturbation theory by Feynman graph
expansions. In particular in lowest order the relation between the bare and
the renomalized mass (\ref{mass}) is given by Figure 1 (a). It had already
been calculated in \cite{KW} and yields
\[
m^{2}=\alpha \left( 1-\frac{1}{6}\left( \frac{\beta ^{2}}{8}\right)
^{2}+O(\beta ^{6})\right)
\]
which agrees with the exact formula above. Similarly we check the quantum
corrections of the trace of the energy momentum tensor (\ref{T}) by
calculating the Feynman graph of Figure 1 (b) with the result again taken
from \cite{KW} as
\[
\langle \,p\,|\,\mathcal{N}\cos \beta \varphi -1\,|\,p\,\rangle =-\beta
^{2}\left( 1+\frac{\beta ^{2}}{8\pi }\right) +O(\beta ^{6}).
\]
This again agrees with the exact formula above since the normalization given
by eq.~(\ref{J}) implies $\langle \,p\,|\,T_{~\mu }^{\mu }|\,p\,\rangle
=2m^{2}$. All other equations have also been checked in perturbation theory
\cite{BK}.

\subparagraph{Acknowledgments:}

We thank J. Balog, S.A. Bulgadaev, R. Flume, A. Fring, R.H. Poghossian, F.A.
Smirnov, R. Schrader, B. Schroer and Al.B. Zamolodchikov for discussions.
One of authors (H.B.) was supported by DFG, Sonderforschungsbereich 288
`Differentialgeometrie und Quantenphysik' and partially by INTAS grant
96-524.


\begin{thebibliography}{99}
\bibitem{Co}  S. Coleman, \emph{Phys. Rev.} {D11} (1975) 2088.

\bibitem{BK}  H. Babujian and M. Karowski, \emph{Exact Form Factors in
Integrable Quantum Field Theories: the Sine-Gordon Model II}, to be
published.

\bibitem{K2}  M. Karowski, \emph{The bootstrap program for 1+1 dimensional
field theoretic models with soliton behavior}, in `Field theoretic methods
in particle physics', ed. W. R\"{u}hl, (Plenum Pub. Co., New York ,1980).

\bibitem{KT}  M. Karowski and H.J. Thun, \emph{Nucl. Phys.} \textbf{B130}
(1977) 295.

\bibitem{KW}  M. Karowski and P. Weisz, \emph{Nucl. Phys.} \textbf{B139}
(1978) 445.

\bibitem{Sch}  R. Schrader, \emph{Fortschritte der Physik, }\textbf{22}
(1974) 611-631.

\bibitem{Fr}  J. Fr\"{o}hlich, in ''Renormalization Theory'', ed. G. Velo et
al. (Reidel, 1976) 371.

\bibitem{ST}  B. Schroer and T. Truong, \emph{Phys. Rev.} \textbf{15} (1977)
1684.

\bibitem{Ko}  V. E. Korepin, \emph{Commun. Math. Phys.} \textbf{76} (1980)
165.

\bibitem{K3}  M. Karowski,\emph{\ Nucl. Phys.} \textbf{B300} [FS22] (1988)
473; \newline
---, \emph{Yang-Baxter algebra - Bethe ansatz - conformal quantum field
theories - quantum groups}, in `Quantum Groups', Lecture Notes in Physics,
Springer (1990) p. 183.

\bibitem{LeC}  A. LeClair, \emph{Phys. Lett. }\textbf{B230} (1989) 103-107.

\bibitem{Sm1}  F.A. Smirnov, \emph{Commun. Math. Phys. }\textbf{131} (1990)
157-178.

\bibitem{Z}  A.B. Zamolodchikov, \emph{JETP Lett. }\textbf{43} (1986) 730;
\emph{Sov. J, Nucl. Phys. }\textbf{46} (1987) 1090.

\bibitem{Ca}  J.L. Cardy,\emph{\ Phys. Rev. Lett. }\textbf{60} (1988) 2709.

\bibitem{KS}  J.M. Kosterlitz and J.P. Thouless, \emph{Journ. Phys. }\textbf{%
C6} (1973) 118.

\bibitem{JKKN}  J. Jose, L. Kadanoff, S. Kirkpatrick and D. Nelson,\emph{\
Phys. Rev. }\textbf{B16} (1977) 1217.

\bibitem{Wi}  P.B. Wiegmann, \emph{Journ. Phys. }\textbf{C11} (1978) 1583.

\bibitem{Za}  Al.B. Zamolodchikov, \emph{Int. Journ. of Mod. Phys.}\textbf{\
A10} (1995) 1125-1150.

\bibitem{BFKZ}  H. Babujian, A. Fring, M. Karowski and A. Zapletal, \emph{%
Nucl. Phys.} \textbf{B538} [FS] (1999) 535-586.

\bibitem{Sm}  F.A. Smirnov, \emph{Form Factors in Completely Integrable
Models of Quantum Field Theory, Adv. Series in Math. Phys.} \textbf{14},
World Scientific 1992.

\bibitem{FMS}  A. Fring, G. Mussardo and P. Simonetti, \emph{Nucl. Phys.}%
\textbf{\ B393} (1993) 413.

\bibitem{KM}  A. Koubeck and G. Mussardo, \emph{Phys. Lett. }\textbf{B311}
(1993) 193.

\bibitem{MS}  G. Mussardo and P. Simonetti, \emph{Int. J. Mod. Phys.}\textbf{%
\ A9} (1994) 3307-3338

\bibitem{BL}  V. Brazhnikov and S. Lukyanov, $\emph{Nucl.Phys.}$\textbf{\
B512} (1998) 616-636.
\end{thebibliography}
\end{document}